\documentclass[journal]{IEEEtran}
\usepackage{flushend}

\IEEEoverridecommandlockouts
\usepackage{cite}
\usepackage{amsmath,amssymb,amsfonts}
\usepackage{algorithm,algpseudocode}
\algnewcommand{\TRUE}{\textbf{true}}
\algnewcommand{\FALSE}{\textbf{false}}
\algnewcommand\algorithmicforeach{\textbf{for each}}
\algdef{S}[FOR]{ForEach}[1]{\algorithmicforeach\ #1\ \algorithmicdo}
\usepackage{graphicx}
\usepackage{textcomp}
\usepackage{xcolor}

\usepackage[T1]{fontenc}
\usepackage{booktabs}
\usepackage{multirow}
\usepackage{subfig}

\usepackage{csquotes}
\newcommand\tab[1][1cm]{\hspace*{#1}}
\def\BibTeX{{\rm B\kern-.05em{\sc i\kern-.025em b}\kern-.08em
    T\kern-.1667em\lower.7ex\hbox{E}\kern-.125emX}}
\begin{document}

\title{FedAT: Federated Adversarial Training for Distributed Insider Threat Detection}
\author{\IEEEauthorblockN{R G Gayathri, Atul Sajjanhar, Md Palash Uddin, Yong Xiang } \\
\IEEEauthorblockA{\textit{School of Information Technology} \\
\textit{Deakin University}\\
Geelong, VIC 3217, Australia \\
\{gradhabaigopina, atul.sajjanhar, m.uddin, yong.xiang\}@deakin.edu.au}
}

\maketitle

\begin{abstract}
Insider threats usually occur from within the workplace, where the attacker is an entity closely associated with the organization. The sequence of actions the entities take on the resources to which they have access rights allows us to identify the insiders. Insider Threat Detection (ITD) using Machine Learning (ML)-based approaches gained attention in the last few years. However, most techniques employed centralized ML methods to perform such an ITD. Organizations operating from multiple locations cannot contribute to the centralized models as the data is generated from various locations. In particular, the user behavior data, which is the primary source of ITD, cannot be shared among the locations due to privacy concerns. Additionally, the data distributed across various locations result in extreme class imbalance due to the rarity of attacks. Federated Learning (FL), a distributed data modeling paradigm, gained much interest recently. However, FL-enabled ITD is not yet explored, and it still needs research to study the significant issues of its implementation in practical settings. As such, our work investigates an FL-enabled multiclass ITD paradigm that considers non-Independent and Identically Distributed (non-IID) data distribution to detect insider threats from different locations (clients) of an organization. Specifically, we propose a Federated Adversarial Training (FedAT) approach using a generative model to alleviate the extreme data skewness arising from the non-IID data distribution among the clients. Besides, we propose to utilize a Self-normalized Neural Network-based Multi-Layer Perceptron (SNN-MLP) model to improve ITD. We perform comprehensive experiments and compare the results with the benchmarks to manifest the enhanced performance of the proposed FedAT-driven ITD scheme.

\end{abstract}	
   	
\begin{IEEEkeywords}
adversarial learning, deep learning, federated learning, insider threat, non-IID 
\end{IEEEkeywords}

\section{Introduction}
Insider threats are attacks that originate within the organization from entities that are directly or indirectly associated with the resources of the organization. Though these types of attacks often happen unintentionally, the consequences can be very critical \cite{insider_survey}. Due to the nature of the attack, it will be mostly left unnoticed. \textcolor{black}{User behavior analysis requires access to resource log files from the organizations. The data utilized in Insider Threat Detection (ITD) are collected from within the organizations, including logging details, web access patterns, email communications, external device usage, file transfer, etc. ITD is usually performed using conventional Machine Learning (ML)-based methods. However, organizations spread across multiple locations require distributed processing of the data so that the data from all locations contribute to the ITD process. The rarity of insider threats results in skewed class distributions. Moreover, sharing the confidential user data distributed across the various locations to the centralized model can cause privacy concerns. The above-mentioned reasons motivate us to leverage Federated Learning (FL) \cite{fed_learning2} for effective ITD for an organization that is geographically located in multiple places.} 

FL was presented as a deep collaborative (parallel and distributed) learning paradigm to overcome the privacy concerns of the centralized ML approaches \cite{fed_learning2}. In FL, the end devices (also known as clients or parties, or users) only communicate partial updates of a global neural network model, which are then compiled by a central entity (aka server, aggregator, or coordinator). Since FL never allows to share of user data with outside parties directly, using it is meant to promote privacy for the users. As such, FL-based schemes are being employed in numerous application domains, such as natural language processing \cite{fl_nlp}, Internet of Things (IoT) \cite{fl_iot}, and health care \cite{fl_healthcare}. 

Several organizations can cooperatively train a neural network (deep) model for ITD with the help of FL without utilizing shared network data. The performance of the ITD model is enhanced while the training dataset is expanded. On the one hand, according to studies, the volume of training data impacts the efficiency of the ML or Deep Learning (DL)-based detection models \cite{dl_data_shortage}. The internal resource logs are highly confidential, and some clients may have fewer data, while the minimal data for each malicious behavior results in weak individual models. Local data collection is not advised in such a case, as the information is very private and sensitive. Each resource has unique user behavior patterns that can be used to train the detection system. However, the model performance would significantly be improved when data from all locations of the organization could be used for training. Therefore, it is impractical to centralize the data for training because of resource constraints, security problems, and privacy considerations. On the other hand, the data residing in multiple locations can have a different sample space in classes, and the heterogeneous datasets stored in the client locations possess a non-Independent and Identically Distributed (non-IID) property. In addition, due to the rarity of insider activities, all the malicious scenarios need not be available in all the locations, leading to class-level data imbalance within the clients. To mitigate these two serious difficulties (clients' data privacy and security associated with different locations, and non-IID data among the clients), we thoughtfully leverage the two cutting-edge technologies i.e., FL and Generative Adversarial Network (GAN) for distributed ITD, which have their own complementing advantages in these regards. To the best of our knowledge, this is the first attempt to reformulate FL and GAN in ITD. The contributions of this work are summarized as follows. 

\begin{itemize}
   	\item We propose an FL-based approach for distributed ITD, which can be used to implement deep collaborative training among several locations (clients) of an organization to maintain the privacy and confidentiality of the employee data. In addition, our developed ITD model deploys a Self-normalized Neural Network-based Multi-Layer Perceptron (SNN-MLP) network for improving detection efficiency. 
 
   	\item We propose to include a GAN-based federated data augmentation strategy to the non-IID data in the clients, which causes class-level data heterogeneity to perform Adversarial Training (AT) for a more robust model that can be shared across organizations. The main goal is to improve the sample diversity and reduce the correlation between data samples, reducing the over-fitting problem in the local clients. 
   	
   	\item We utilize the two versions of CERT CMU benchmark datasets for generating the non-IID data settings and perform comprehensive experiments to compare various federated optimization algorithms, including Federated Averaging (FedAvg) and Federated Proximal (FedProx), for the distributed ITD. 
\end{itemize}

The remainder of this paper is structured as follows. Section \ref{sec_Lit_Review} summarizes the key concepts adopted in this work. Section \ref{sec_methodology} explains  the proposed method in detail. Section \ref{sec_exp_setup} briefs about the data description used for experiments, followed by a performance evaluation of our proposed approach in terms of various evaluation metrics. Section \ref{sec_conclusion} concludes the paper and provides potential future research directions.

\section{Preliminaries}\label{sec_Lit_Review}
This section provides the background of the key concepts (ITD, FL, and AT) adopted toward the motivation of this paper. 

\subsection{Insider Threat Detection in Distributed Setting}
The analysis of insider threats has been researched extensively for many years \cite{insider1, insider2, insider3, insider4}. However, the research community could not significantly contribute to this attack in a distributed setting. Due to the rise in attack frequency and the emergence of potential data analysis methods, it now draws more attention. Through FL, many organizations can train DL models collaboratively in parallel for ITD independently. Various institutions with significant employees and resource access logs, as in domains like healthcare, education, and finance, may be among the FL participants in the FL-driven pipeline for ITD. Implementing FL in this way is particularly crucial as the organizations never agree to share the data for privacy and confidentiality reasons. However, such a potential distributed approach for ITD is not yet studied.  

\subsection{Federated Learning}
FL  is an ML framework that bases data use for ML on data security and privacy \cite{fed_learning2}. Google came up with the idea first. 
A server and a set of clients comprise the fundamental building blocks of the FL paradigm. The server does not gather data, rather than it gathers model parameters only. Each client maintains the local training dataset, while the server manages client participation in training. The training data is stored locally by the client and used to train the local model before uploading its parameters to the server. For the subsequent training round, the server computes the average of the collected parameters and shares them with each client for retraining. Various aggregation methods are used to synchronize the client models with the global weights. One of the popular aggregation methods is FedAvg \cite{fedavg} which performs the arithmetic averaging of the client parameters. Another method refers to as FedProx \cite{fedprox} adds a proximal term to the local objective function of the client and performs the same aggregation as of FedAvg.

\subsubsection{Local Model Update in FL}

The FL strategy seeks to minimize the following overall objective function ($F(\omega)$) in a parallel way, where $L_k(\omega)$ is the $k^{th}$ client's ($k \in [1, K]$) local objective function and $K$ is the total number of clients participating in the training process. 

\begin{equation}
    _{{\omega\in{\mathbb{R}^d}}}^{min} F(\omega)=\sum_{k=1}^{K}\frac{N_k}{N}L_{k}(\omega),
    \label{classical_obj_func}
\end{equation}
where $N_k$ represents the total number of samples of the $k^{th}$ client, $N=\sum_{k=1}^{K}N_k$, and $L_{k}(\omega)$ is calculated using the sample ($x_i$) loss, $l_i$ as follows.
\begin{equation}
    L_{k}(\omega)=\frac{1}{N_k}\sum_{i=1}^{N_k}l_i(\omega, x_i).   \label{classical_loss_calc}
\end{equation}

In this way, as the model's weight ($\omega$) changes, the FL minimizes the weighted average of the clients' local losses ($L_k(\omega)$), hence lowering the cumulative loss function ($F(\omega)$). The two most promising federated optimization algorithms, FedAvg and FedProx, are now presented to demonstrate how to minimize the Eq. (\ref{classical_obj_func}) in our proposed FL-driven ITD setup. FedAvg starts the training process by sending the first global model to a small sample of clients. The retrieved shared global model is then trained in parallel on each client's private data samples, and the trained model is then uploaded to the FL server for aggregation. Until convergence, the training and aggregation procedures are repeated.  However, FedProx alters the local objective of Eq. (\ref{classical_loss_calc}) by inserting a proximal factor, as shown below, and then follows the same processes as FedAvg.

\begin{equation}
    _{{\omega\in{\mathbb{R}^d}}}^{min} L_{k}(\omega^{k}_{t+1}, \omega_t)=L_{k}(\omega^{k}_{t+1})+ \frac{\mu}{2}||\omega^{k}_{t+1}-\omega_t ||^{2},   \label{classical_loss_calc_prox}
\end{equation}
where $\mu$ represents the regularized coefficient.  

\subsubsection{Global Model Aggregation in FL}
During a global round, all locally trained models are uploaded to the server, and then the global model aggregation is accomplished using the traditional arithmetic mean method. The server specifically calculates the aggregate global model ($\omega_{t+1}$) as the weighted average of the local models' weights ($\omega_{t+1}^k$) from the $K$ clients, as shown below.

\begin{equation}
\omega_{t+1}= \sum_{k=1}^{k}\ \frac{N_k}{N} \omega_{t+1}^{k}.
\label{eqn:aggregation}
\end{equation}

This global model is then sent to the clients for retraining using the local model update mechanism. These procedures of training and aggregation are repeated until achieving the targeted convergence of the global model. 

\subsection{Adversarial Training}
Adversarial samples are data samples that an attacker can design to make the model intentionally make mistakes \cite{adv_samples}, which are employed for varying purposes, such as developing robust models, and data augmentation \cite{adv_data_aug}. The work in \cite{adv_samples_1} considered adversarial samples as features where the sensitivity to well-generalizing features in the data directly leads to adversarial vulnerability. GAN \cite{gan} proved to be an efficient method to generate adversarial samples. The two fundamental models in GAN are the generative model (G), and the discriminative model (D). The G model makes an effort to provide examples that fool the D model. When the GAN has finished training, the D model cannot distinguish between samples produced by the G model and actual data. Technically, G takes a uniform random noise $z$ as input and generates synthetic (fake) data samples that closely mimic the original data distribution; it tries to learn the probability $P(X)$ for input $X$. D tries to distinguish between the real and fake data samples; it learns the conditional probability $P(Y|X)$ for input $X$ with class labels $Y$. G is trained to maximize the classification error between the real and the fake data while D is trained to minimize the same. In this way, the GAN training follows a game theory principle of reaching an equilibrium when the generator produces data from the original distribution such that the discriminator fails to identify between real and fake and gives a probability of flipping the coin. The GAN training happens such that G and D converge together. What follows, the below equation depicts the GAN optimization. 

\begin{equation}
    \min_{G}\max_{D}\mathbb{E}_{x\sim p_{\text{data}}(x)}[\log{D(x)}] +  \mathbb{E}_{z\sim p_{\text{z}}(z)}[1 - \log{D(G(z))}].
\end{equation}

Conditional GAN (CGAN) \cite{cgan} is an extended mechanism of the vanilla-GAN. Compared to vanilla-GAN, the condition variable, which is additional information, must be present in the input variables of the generator of the CGAN. Class labels of the actual data are generally used as the conditioning information. ACGAN \cite{acgan}, a variant of CGAN, is capable of controlling the output with additional input. The input to the G model of ACGAN is noise from latent space and the class label from the original dataset to condition the data generation. But the D model of ACGAN considers only the real and synthetic data samples. Like CGAN, the D model of ACGAN predicts whether the given data is real or fake along with the class label. In \cite{fl_adv_learn}, the authors adopted adversarial techniques to the constraints of the federated environment to align the representations learned among many clients. Although the original data can be created via the shared vectors, this effort violates the federated configuration by passing the output vectors of local models to train the discriminators.




\subsection{Motivation}
FL proved its efficiency in numerous application domains including intrusion detection \cite{fl_ids_survey}, natural language processing \cite{fl_nlp}, and health care \cite{fl_healthcare} while it is adopted for distributed anomaly detection in numerous domains, such as industrial control, and IoT field. However, there is no existing ITD research that leverages FL in various scenarios (classes). We observe that insider threats are a classical real-world example of the extreme imbalance in classes within and across the clients in an FL setting. Motivated by the success of solution approaches that adopt FL for distributed training and AT for effective data augmentation, we propose a Federated AT (FedAT) for effective ITD in this paper. 

\section{Proposed Approach}\label{sec_methodology}
\subsection{Approach Overview}
Insiders are entities associated with the organization having access to one or more organizational resources. The activities of an entity associated with the organization  referred to as scenarios (classes) identify the action sequences resulting in the insider threat. Organizations operate from multiple places scattered across different locations. Distributed analysis methods can help build effective threat detection, and FL can be an appropriate choice in this regard. As such, we propose a FedAT-enabled ITD approach that can help work in a real-world distributed setting.  

\begin{figure}[tbph!]
   	\centering
   	\includegraphics[scale=0.2]{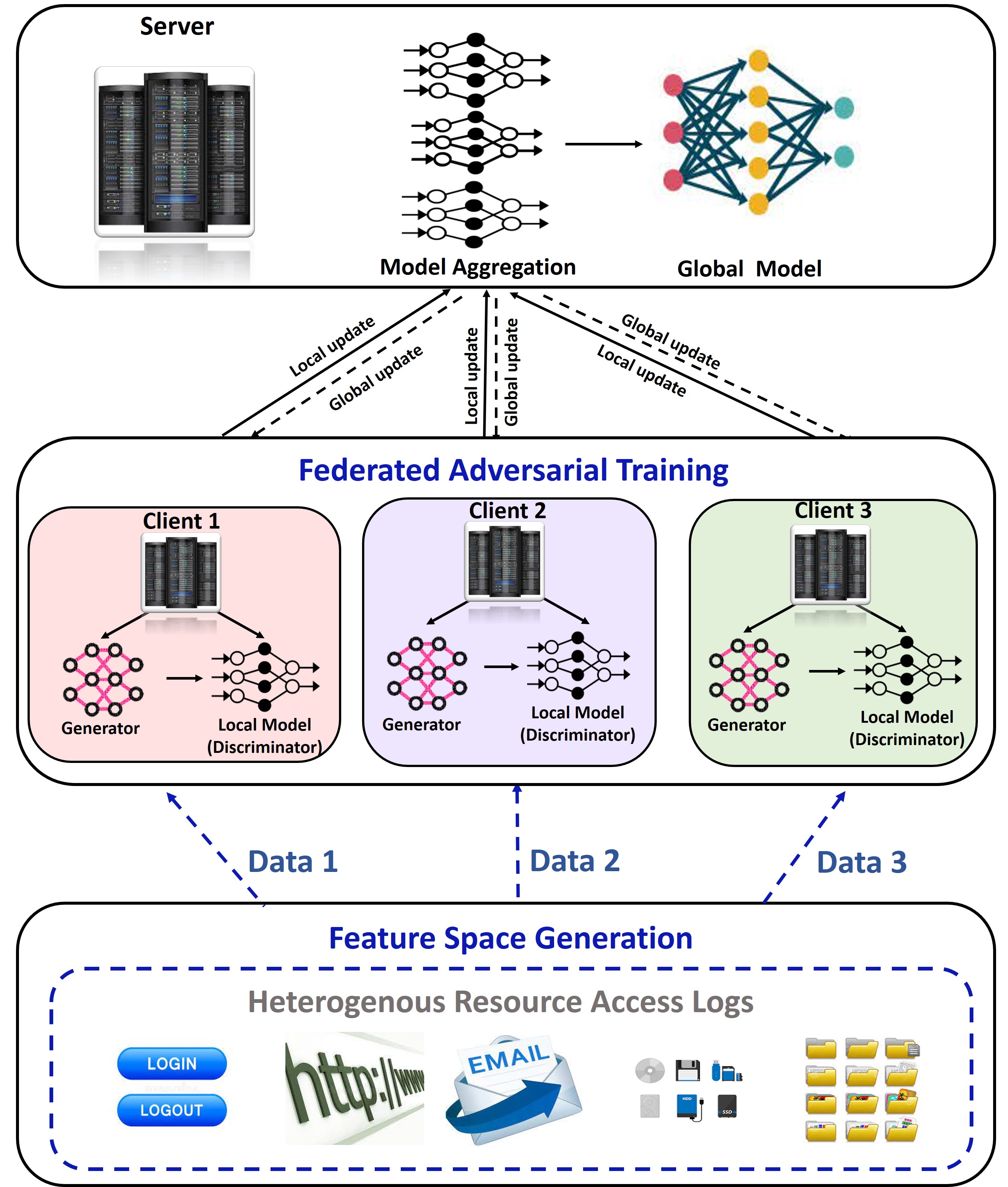}
   	\caption{FedAT-enabled ITD framework. The first step is feature space generation which performs the feature extraction and data preparation. The next step is the FedAT in the local clients, followed by model aggregation in the server.
        }
   	\label{fig:distributed}
\end{figure}

We consider an organizational environment that uses data from multiple client locations, where the clients do not contain all the scenarios of data samples. In this way, the data follows a non-IID setting that results in class-level imbalance across the clients. In addition, due to the rarity of the insider threat and the fewer chances of noticing suspicious actions, each client has a class imbalance among the malicious and non-malicious insider scenarios. In particular, let us consider a real-world setting, where there are $K$ clients and an aggregation server $S$. In addition, let each client have its private dataset. Then, Fig. \ref{fig:distributed} provides an overall architecture being followed in the proposed FedAT-driven ITD framework. 

\textcolor{black}{As illustrated in Fig. \ref{fig:distributed}, the overall process is split into the following sequential stages: \textcolor{black}{(i) distributed feature space generation; and (ii) and FedAT followed by the multiclass classification using SNN-MLP}. Particularly, in the first stage, the clients perform the feature generation and pre-processing of the heterogeneous resource access files in the respective locations, as shown in Fig. \ref{fig:distributed}. The labeled data in each client results in a drastic class imbalance that hinders the performance of any ML algorithm. The private dataset of the client is the input to the second stage of the FedAT process, including the synthetic data generation and the AT that performs the multiclass classification for insider threat detection. Once the local training (FedAT) is complete, the local model updates are uploaded to the server. The server then performs the model aggregation, and the updated model is broadcasted again to the clients for retraining with FedAT.}

\subsection{Distributed Feature Space Generation} \label{sec_ftr_extr}
\begin{figure}[hbt]
   	\centering
   	\includegraphics[scale=0.21]{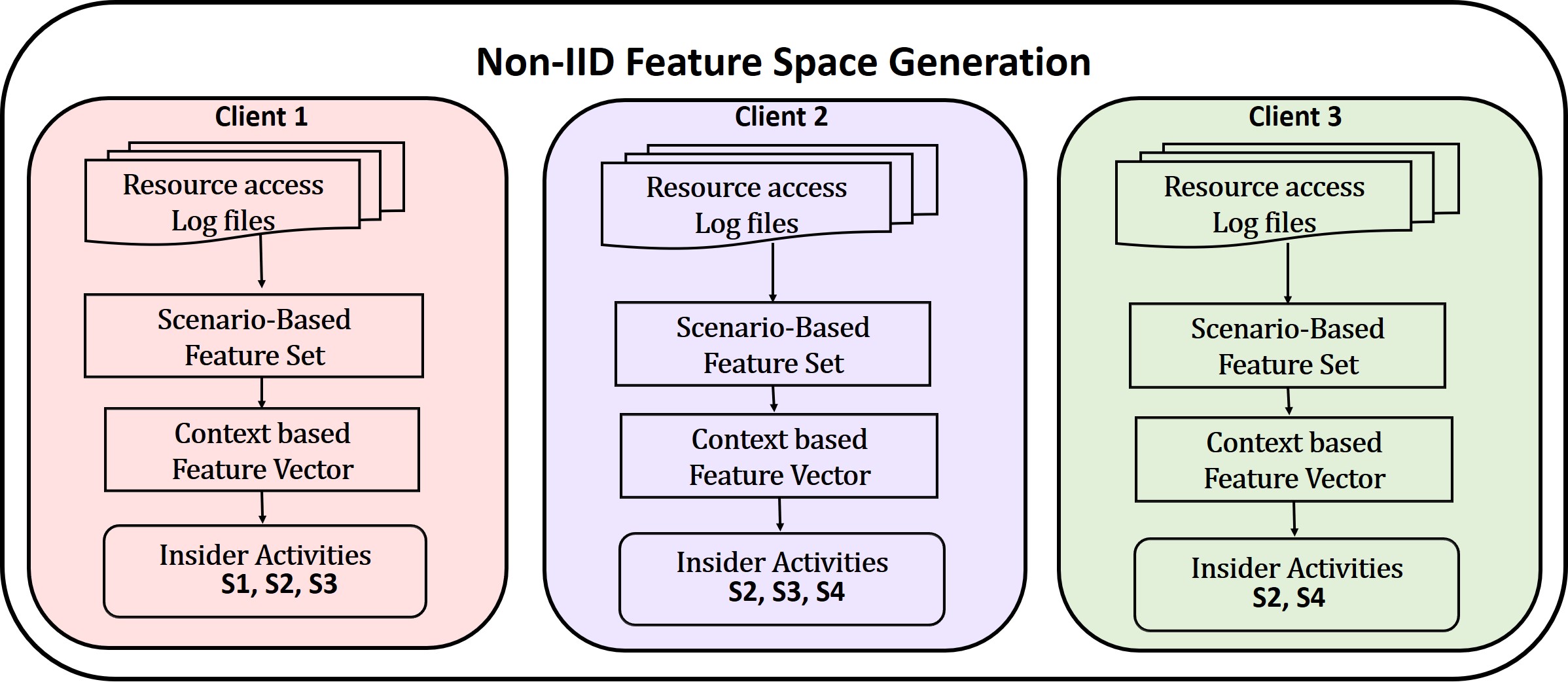}
   	\caption{Non-IID insider threat data generation. Each client performs the feature space generation independently and creates a context-based feature vector using the log files available at the respective site. Classes are labeled after the scenarios that result in an insider threat.
        }
   	\label{fig:ftrgen}
\end{figure}

Implementing the cooperative training model with other organizations working in the same field is more important for obtaining a superior DL-based ITD model. The clients of FL are these organizations. The server can be in the cloud. Once FL is completely implemented, each client will automatically connect to the server. Depending on the circumstances, the client can also terminate FL whenever appropriate. All clients do not have to engage in each cycle of learning simultaneously with FL. The input for the ITD is the behavior of the users. The user behavior is extracted from the heterogeneous resource access files obtained from within the organizations. The activities include the logon-logoff details, the email communication, the web access pattern, the external device, file usage, etc. The user behavior pattern specifies the malicious non-malicious action sequence. Any deviation from normal behavior results in suspicious activity. In this work, we follow the feature extraction adopted in \cite{insider_cgan_ours} and reformulate in our FL setting, while Fig. \ref{fig:ftrgen} provides an overview of the non-IID feature generation followed in this work of distributed setting. As seen in Fig. \ref{fig:ftrgen}, each client performs private feature extraction from the heterogeneous files to generate its private ITD dataset. The set of employee activities that result in a malicious action is referred to as a scenario (class). Each client will have different scenarios present in its dataset, resulting in heterogeneity in the class distribution in the client datasets, which leads the local datasets to become non-IID.

\subsection{FedAT}
The privacy concerns of the employee data pose a hindrance to effective ITD. FL is among the most effective methods for safeguarding data privacy in ML models. The client's data is kept secure by simply providing the model's updated weights and not the actual data. This strategy, which retrains each client's model using baseline data, addresses the problem of non-IID data. As such, we propose an FL-driven method for distributed ITD. As per the proposed architecture (Fig. \ref{fig:distributed}), there are $K$ clients distributed across various locations. To avoid sharing private data, the clients perform local model building at the client site, and the model parameters are shared with the server. The server conducts model aggregation using promising optimization methods like FedAvg and FedProx. However, the non-IID data residing in the client locations have a drastic class imbalance, which affects the performance of the federated training; hence raises the need to address the class-level data scarcity in the clients, which inspires us to augment the minority samples across the client locations. In addition, when the training data is scarce, the traditional FL models tend to perform poorly and produce higher false alarm rates. To address these issues, we propose to adopt the GAN network to synthesize the minority data samples in the FL setting for effective ITD.

Inspired by the GAN-based FL works, such as Multiple Discriminators-based GAN (MDGAN) \cite{mdgan} and FL-GAN \cite{priv_fl_gan} models, which are proposed for protecting the privacy and creating synthetic data, we thoughtfully leverage GAN in this work. In MDGAN, there are multiple D models at the client locations and just one G model at the server. The data from the G model is shared among various clients by distributing the data. In the FL-GAN architecture, the workers (clients) also have G models and D models in addition to the server. FL-GAN distributes data and shares the updates averaged by the server with the clients. 
However, in our FedAT-driven ITD framework, each client has only one G model to lightweight the client-side computation, which is used to create synthetic insider samples that resemble those produced by standard GAN modeling. The classifier (C) in the local client is used as the D model. As such, this is the first work to leverage both FL and GAN architecture to construct multiclass classification-based distributed ITD.

\begin{figure}[tbph!]
   	\centering
   	\includegraphics[scale=0.2]{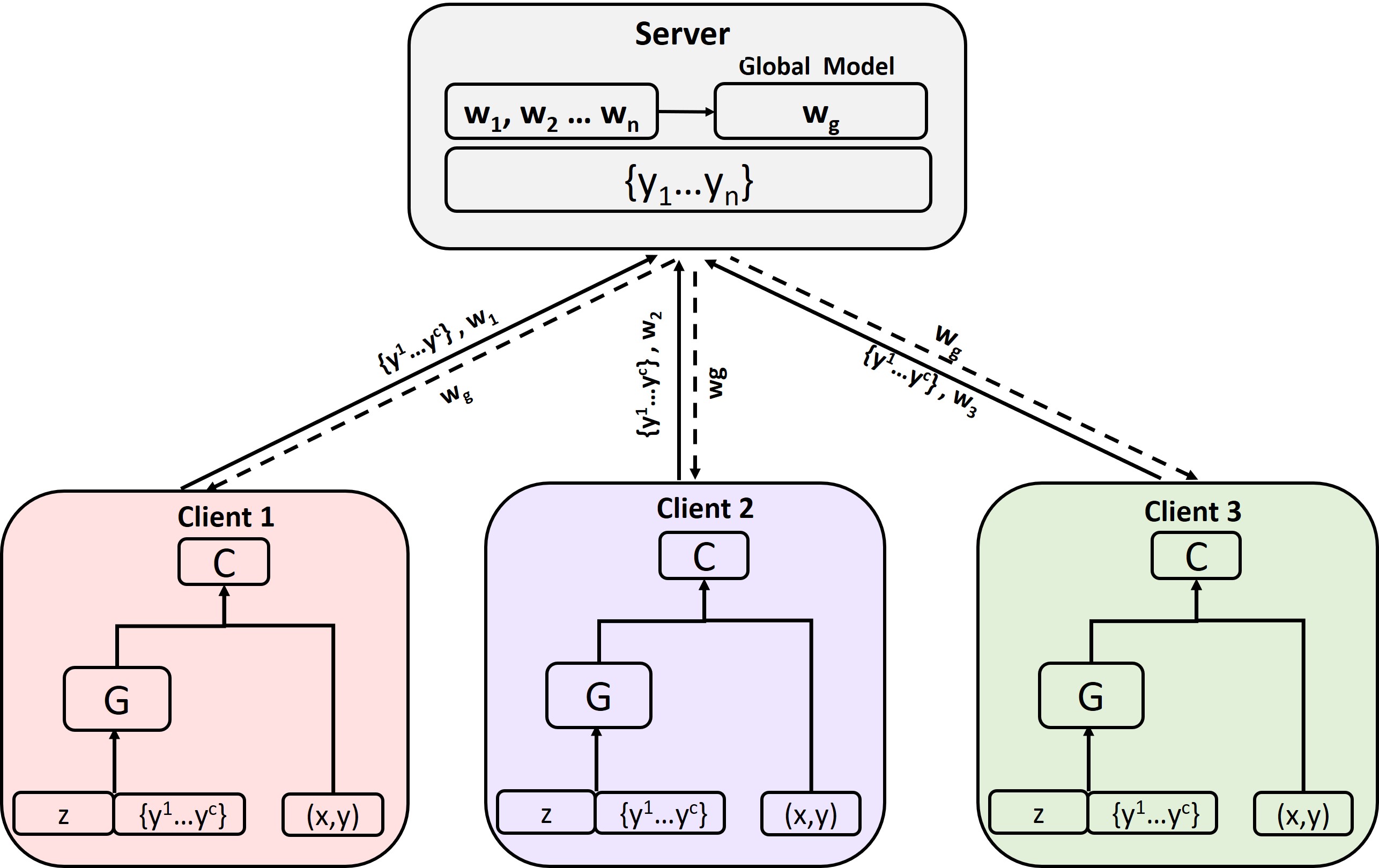}
   	\caption[FL]{FL with local GAN augmentation for ITD. As the first step in FedAT, the GAN training is employed by the G model and the C model. Later, the C model is used for multiclass classification. Finally, the gradients from the C model and the auxiliary information (the class labels) are shared with the server. 
        }
   	\label{fig:fedlearning}
\end{figure}

\begin{algorithm}
   	\caption{FL: $K$ is the number of clients; $B$ is the local batch size; $E$ is the number of local epochs; $T$ is the number of communication rounds; $\eta$ is the learning rate; \textcolor{black}{$c$ is the number of classes present in each client.}}
   	\label{algo_fed_learn}
   	\begin{algorithmic}[1]
   		\Procedure{ServerUpdate:}{}
   		\State Initialize weights,  ${\omega }_{t}$, \textcolor{black}{auxiliary information ($y$)}
   		\ForEach{communication round, $t = 1, \dots, T$}
   			\ForEach{client, $k = 1, \dots, K$ \textbf{in parallel}}
   				\State ${\omega }_{t+1}^{k}, y_{t+1}^k \gets\Call{ClientUpdate}{k,w\textsubscript{t},y}$ 
                    \newline  \tab \Comment{GAN-based local model training}
   			\EndFor
   			\State $\omega_{t+1} \leftarrow \sum_{k=1}^{K}\frac{n^{k}}{n} \omega_{t+1}^{k}$
                    \newline  \tab \Comment{Global model aggegation}
   			\textcolor{black}{\State $y_{t+1} \leftarrow \bigcup\limits_{k=1}^{K} y_{t+1}^k$}
   		\EndFor
   		\EndProcedure
   		\newline
   		\Procedure{ClientUpdate}{$k,w\textsubscript{t},y$}
   		\ForEach{local epoch, $i = 1, \dots, E$}
		   \State Set $\omega_k=\omega_t$
		   \State Sample noise from $N(0,\sigma^2)$
              \textcolor{black}{\State AT - Update the \textit{G model} and the \textit{C model}}
		   \State Freeze the \textit{G model}
		   \State $ x_{adv} \leftarrow G(y^c) $
		   \State $ x_{merged} \leftarrow x_{train} + x_{adv}$
		   \State Split $x_{merged}$ into batches of size $B$
		   \ForEach {batch, $b \in B$}
   		   	 	\State $\omega_{k} \leftarrow \omega_k - \eta L(\omega_k,b)$
   		   	 	\textcolor{black}{\State $y_{k} \leftarrow \left \{y_1 , y_2 ... y_c \right \}$}	
		   \EndFor
  		 \EndFor
  		 \State return $\omega_{k}$ and $y_{k}$ to the server
   		\EndProcedure
   	\end{algorithmic}
\end{algorithm}

\subsubsection{Federated Data Augmentation using AT} 
AT using synthetic data samples has certain advantages in FL for distributed ITD. When the training data is mixed with other synthetic data that computes different parameters, launching a member inference and inversion attack to get the real dataset or a sample of the dataset becomes difficult. In addition, the order of the parameter transfer and aggregation techniques affects the model's accuracy. With these arguments, we investigate ITD models built using GANs. We design a GAN-based FL setting for ITD using multiclass classification. The approach uses an ACGAN in each client. The client performs two processes: (i) ACGAN training to generate synthetic data, which provides the model to generate data for specific scenarios (classes), and (ii) multiclass classification using our designed SNN-MLP model. Fig. \ref{fig:fedlearning} shows that each client has a G model and a Classifier (C) as the D model. The G model generates the synthetic data samples from a Gaussian distribution $z$ and the class labels $\{y_1..y_c\}$ of the original samples present in that client. The D (C) model in our ACGAN architecture is a classifier that takes the real and generated data as inputs and provides a classification outcome. Therefore, AT happens to the local clients. The operations of the D model (aka the C model) are performed by the SNN-MLP. The label information from each client is passed on to the server as auxiliary information. This information is used by the other clients in the succeeding training rounds, which helps fine-tune the other local C models in the learning process, resulting in turn in the development of a robust model. Once the GAN training is accomplished, the G model is used to generate synthetic samples for the minority class samples present in the client. These are the adversarial samples merged with the original ones to perform the AT. The classifier performs the local training and shares the local updates and the auxiliary information with the global server. The auxiliary information is the scenarios (class labels) present in each client. The local parameters are aggregated in the server, and the updates are disseminated to the clients. In this way, the clients can access the scenarios available across all the clients, which helps gradually refine the GAN model without sharing the original data. In addition, the classifier takes two-folded responsibilities where it serves as the D and the local C models, thereby reducing the complexity of computation in each client. To summarize, Algorithm \ref{algo_fed_learn} provides the detailed implementing pseudocode of the proposed FedAT process.

\subsubsection{Multiclass Classification using SNN-MLP} 
\begin{figure}[!hbt]
   	\centering
   	\subfloat[b][SNN-MLP.]{\includegraphics[scale=0.28]
   		{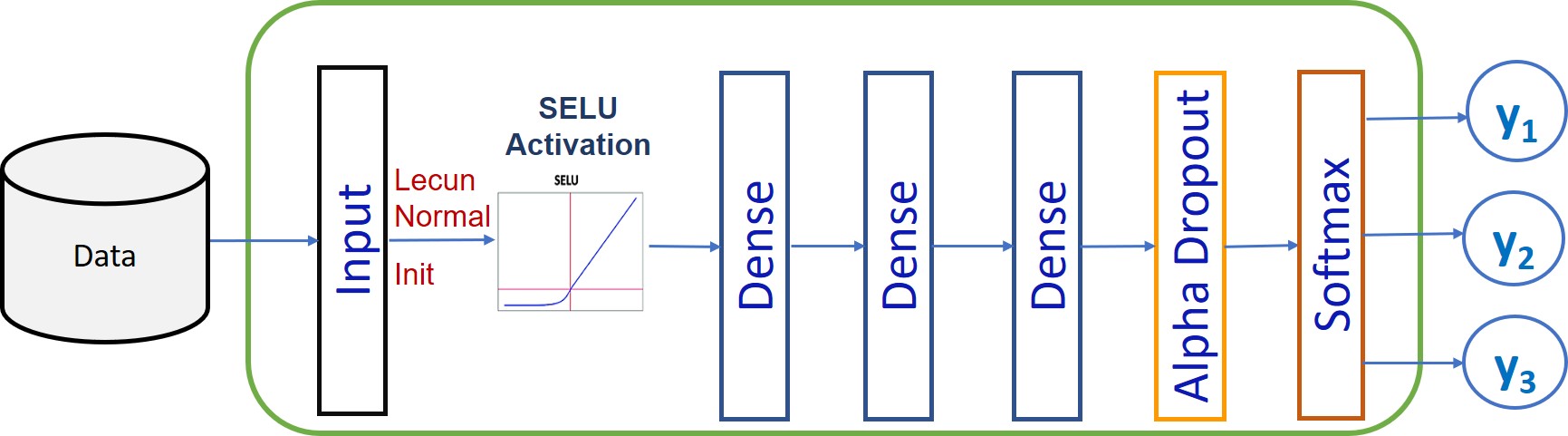}
   		\label{fig:snnmlp}}
   	\newline
   	\subfloat[b][Traditional MLP.]{\includegraphics[scale=0.28]
   		{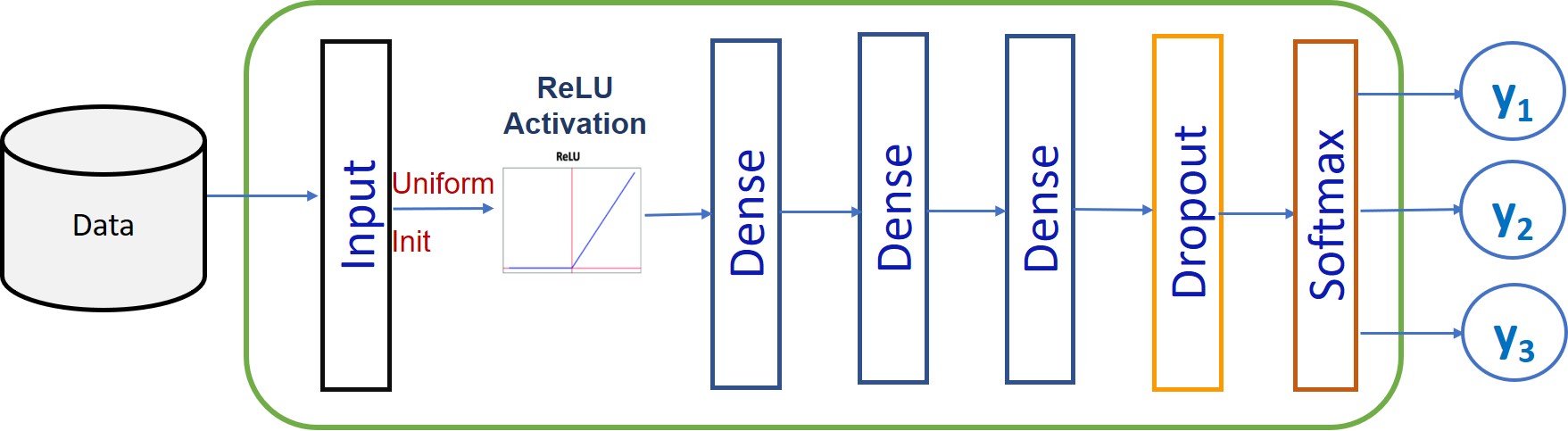}
   		\label{fig:mlp}}
   	\caption{Network architectures of the SNN-MLP vs the classical MLP. (a) depicts the proposed SNN-MLP model, where SELU activation and AlphaDropout are adopted. (b) illustrates the adoption of the ReLU activation and the standard Dropout in the classical MLP model.
        }
\end{figure}

The classical MLP network equipped with the Rectified Linear Unit (ReLU) activation suffers from the vanishing gradient issue as it clips the negative values to 0. It is preferred to have an activation that can contain positive and negative values to control the mean, aids in reducing the variance, and slope greater than one to increase the variance when it is too small. As such, our proposed classifier network is designed with the SNN-based MLP \cite{snn}, called SNN-MLP, where the Scaled Exponential Linear Unit (SELU) function is employed to fulfill the aforesaid properties of the desired activation function, defined in Eq. (\ref{eq_selu}).

\begin{equation}
   	\label{eq_selu}
   	SELU(x)=
   	\left\{\begin{matrix}
   		x, & if  x > 0\\
   		\lambda\alpha (e^{x} - 1),  & otherwise
   	\end{matrix}\right.,
\end{equation}
where $x$ is the input, $\alpha$ and $\lambda$ are the hyperparameters, and $e$ represents the exponent operation. As such, we apply a SELU layer to the subsequent dense layers to further boost the model resistance to the effects of vulnerability due to the invariance \cite{snn_vuln}. The G model in each client is, therefore, a two-layer perceptron with the $tanh$ activation function. A softmax function is finally used in the last layer for the multiclass classification. To this end, the network architectures of the classical MLP and the SNN-MLP realized in this work are illustrated in Fig \ref{fig:snnmlp}, and Fig \ref{fig:mlp}, respectively. Notice that the AlphaDropout layer after the SELU activation in the SNN-MLP randomly determines the  values of neurons rather than the zero value that results from a standard dropout of the classical MLP, maintaining the mean and variance at 0, and 1, respectively. At last, a Lecun Uniform Initializer (LUI) \cite{lecun_layer} is used to initialize the weights of the SNN-MLP network.

\section{Experiment and Result Analysis}\label{sec_exp_setup}
In this section, we evaluate the performance of the proposed FedAT-driven distributed ITD paradigm. We first provide the dataset details followed by the training and testing performance evaluation. We aim to assess the effectiveness of the proposed FedAT scheme using different federated optimization methods (FedAvg and FedProx) and compare the results against the benchmark results from the centralized ML approach.

\subsection{Dataset Description}
\begin{figure}[!hbt]
   	\centering
   	\subfloat[][Non-IID data distribution for the CERT v4.2 dataset with $K=3$.]{\includegraphics[width=0.4\textwidth]
   		{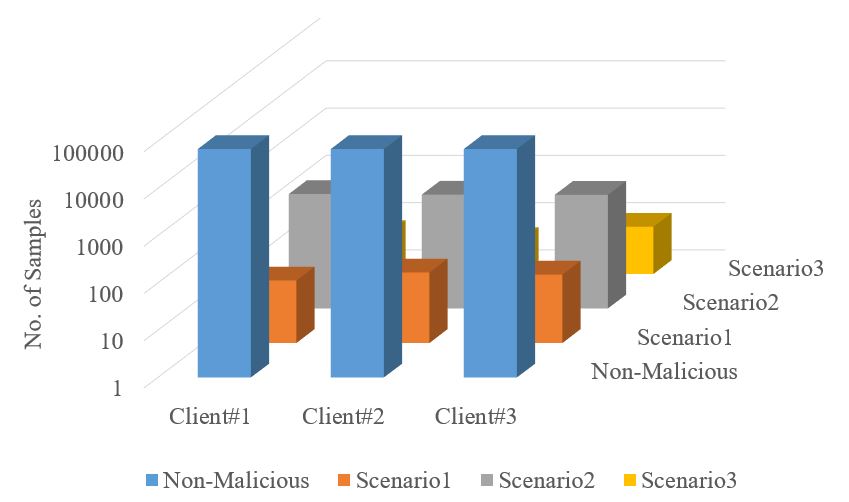}
   		\label{fig:noniidcert42}}
   	\newline
   	\subfloat[][Non-IID data distribution for the CERT v5.2 dataset with $K=5$.]{\includegraphics[width=0.4\textwidth]
   		{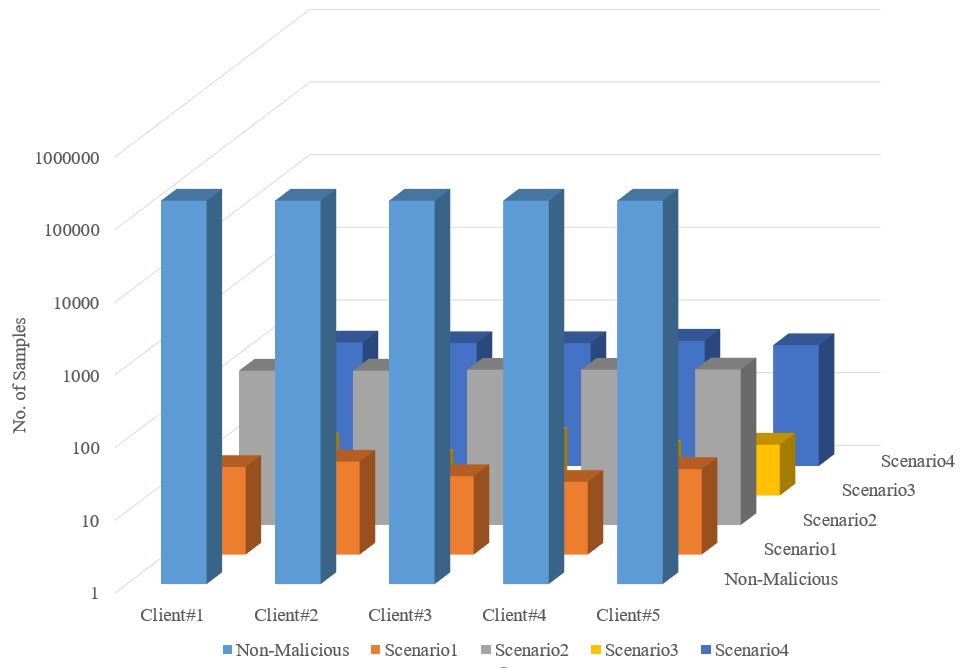}
   		\label{fig:noniidcert52}}
   	\caption{Non-IID data distribution for the CERT datasets. }  
   	\label{fig:noniiddata} 
\end{figure}

The lack of real-world ITD data significantly hinders the experimentation. Usual practice demands using either data gathered from actual user data or artificially created data. The majority of malicious insiders are usually employees with access rights in an organization. Direct tracking and observation of user behavior and employee actions constitute data collection, resulting in concerns about confidentiality and privacy in the workplace. Researchers are thus forced to progress with synthetically generated data in such situations. What follows, we use the CMU CERT dataset \cite{cmu}, the widely accepted synthetic data for ITD, for our experimental analysis. Although there are different CERT dataset versions, we first use CERT v4.2, which is frequently utilized in the ITD literature because it has the most insider cases organized into three scenarios and the data includes 70 insiders from 1000 users over 500 days. In addition, we use another version, CERT v5.2, having 2000 users and 30 insiders from five scenarios. Due to the scarcity of the data, we choose two options with the number of clients to generate the non-IID data; $K = \{3,5\}$. The non-IID data generation is undertaken to divide the classes among the various clients. Each client owns a specific set of insider scenarios as class labels, as shown in Fig \ref{fig:ftrgen}. Besides, Fig \ref{fig:noniiddata} gives a clear overview of the severe non-IID data distribution for the CMU CERT v4.2 dataset with three clients, and the CMU CERT v5.2 dataset with five clients.

\subsection{Benchmark Setup with Centralized ML} 
We first use the centralized ML paradigm (where all data from every client are stored in a single device to train the global model) to evaluate the performance of the MLP and SNN-MLP networks on the CERT datasets. 
The experiments are accomplished on the original data setups, which have an extremely imbalanced scenario, and the AT-driven augmented data settings to compare the performance of the original and the augmented datasets. We consider these results as the benchmarks for the performance analysis as no works exist using the FL approach for ITD. Since the data are originally imbalanced, we consider the Precision ($P$), Recall ($R$), and F-score ($F$) metrics for evaluation purposes. The results for the centralized ML model with and without AT are provided in Table \ref{tab:benchmark}. As seen in Table \ref{tab:benchmark}, MLP is not able to perform well on the original data setups, whereas there is a significant improvement in the performance on the augmented datasets, generated via GAN-based AT. The results of the proposed SNN-MLP, therefore, give a remarkably enhanced performance for AT than training using the original data on both datasets. Though AT with SNN-MLP gives an increased performance of 0.02 than MLP on the CERT v4.2 dataset, there is evidence of a more increase i.e., 0.18 in the performance on CERT v5.2, which has originally more class imbalance.

\begin{table*}[hbt]
   	\centering
   	\caption{Performance analysis in the centralized learning  setup.}
   	\label{tab:benchmark}
   	\begin{tabular}{ccllllll}
   		\hline
   		\multirow{2}{*}{{Model}}   & 
   		\multirow{2}{*}{{Data}} & 
   		\multicolumn{3}{c}{{CERT v4.2}}                                                                           & \multicolumn{3}{c}{{CERT   v5.2}}                                                                           \\ \cline{3-8} 
   		&                                  & \multicolumn{1}{c}{{$P$}} & \multicolumn{1}{c}{{$R$}} & 
   		\multicolumn{1}{c}{{$F$}} & \multicolumn{1}{c}{{$P$}} & \multicolumn{1}{c}{{$R$}} & 
   		\multicolumn{1}{c}{{$F$}} \\ \hline
   		\multirow{2}{*}{{MLP}}     & {Original}                & 0.2750                                 & 0.7820                               & 0.2830                                & 0.3085                                & 0.4133                              & 0.3366                               \\ 
   		& {AT}                      & 0.7582                                & 0.6681                              & 0.6777                               & 0.6360                                 & 0.7793                              & 0.6530                                \\ \hline
   		\multirow{2}{*}{{SNN-MLP}} & {Original}                & 0.2765                                & 0.5695                              & 0.2893                               & 0.2050                                 & 0.6986                              & 0.1717                               \\ 
   		& {AT}                      & 0.7790                                 & 0.7890                               & 0.7830                                & 0.8150                                 & 0.7480                               & 0.7790                                \\ \hline
   	\end{tabular}
\end{table*}

\subsection{Classical FL vs FedAT}
In this section, we perform experiments on the classical FL environment and the setup explained in our proposed FedAT-driven ITD architecture, illustrated in Fig \ref{fig:distributed}. 
As the IID setting is similar to the centralized ML approach for this dataset and the application, we use the non-IID data setting for these experiments, as mentioned in Section \ref{sec_ftr_extr}. In particular, the CERT datasets are divided among various clients $\{1,2..K\}$, and the class-level heterogeneity among the clients is ensured by splitting the data so that each client has one or more scenarios (classes) in the local private data. Splitting the data across multiple clients also reduces each client's private data size. As the number of malicious samples is minimal compared to the non-malicious (normal) data samples, we split the data among fewer clients, $K=\{3,5\}$. However, the proposed method is scalable to any number of clients.

\begin{figure*}[!hbtp]
   	\centering
   	\subfloat[][Classical FL with MLP using $K=3, B=128, E=1,$ and $ T=60$.]{\includegraphics[width=0.45\textwidth]
   		{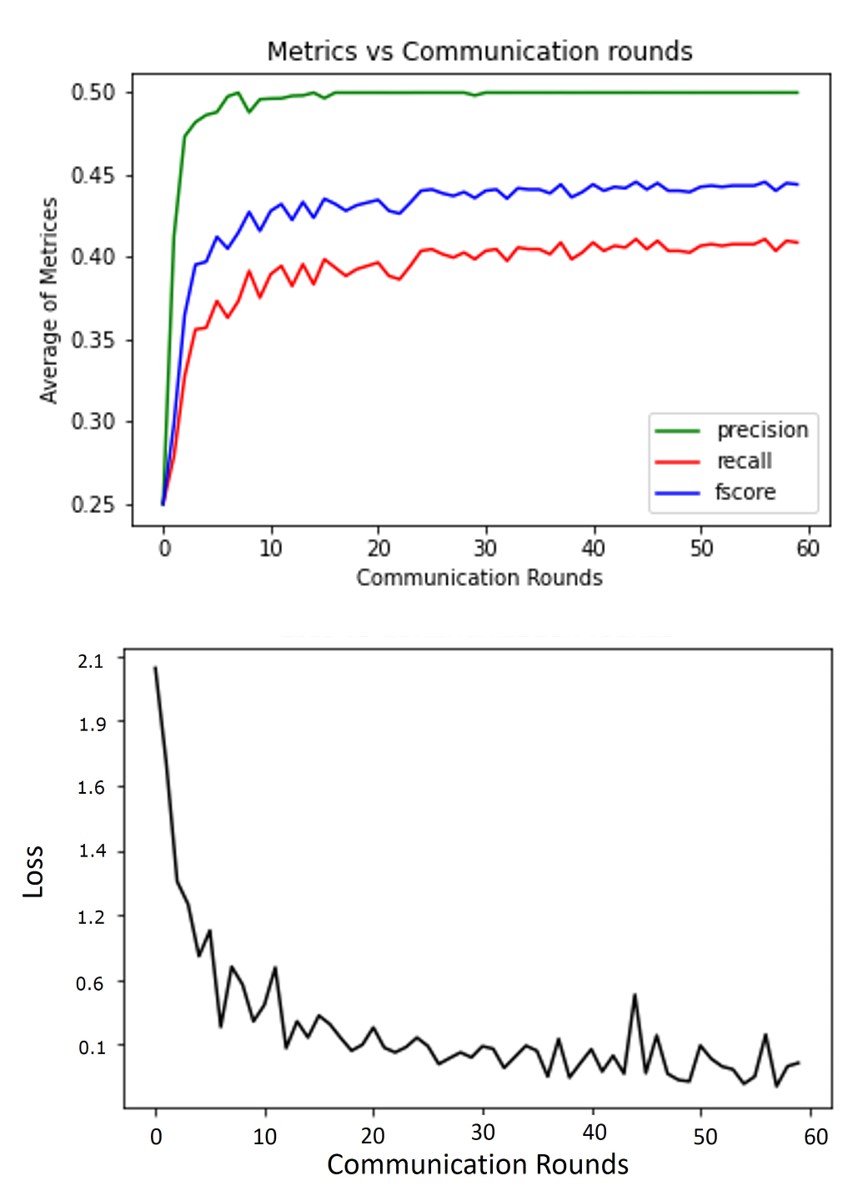}
   		\label{fig:mlp_metrics}}
    \quad 
    \quad 
   	\subfloat[][FedAT with SNN-MLP using $K=3, B=128, E=1,$ and $ T=60$.]{\includegraphics[width=0.48\textwidth]
   		{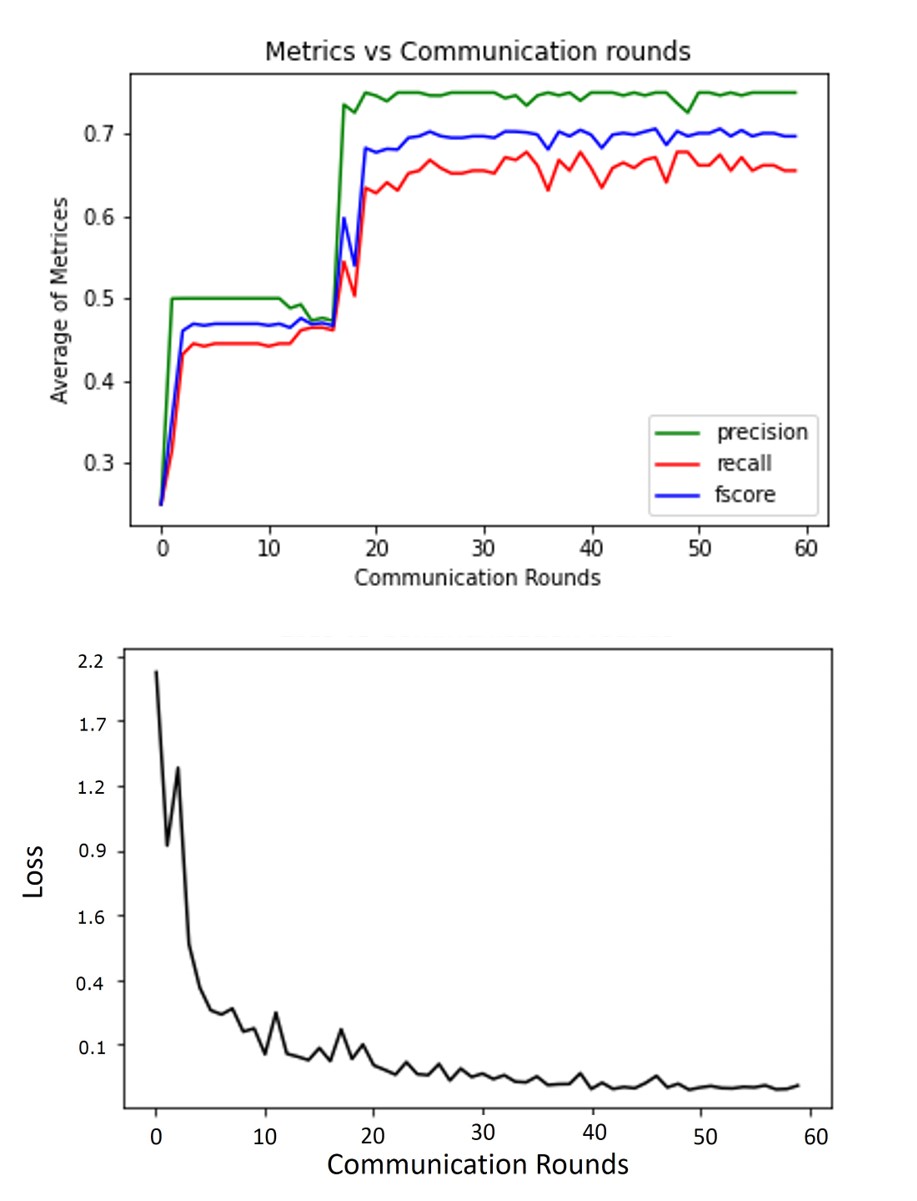}
   		\label{fig:snnmlp_metrics}}
   	\caption{Round-by-round testing performance and training loss. }  
   	\label{fig:metrics_loss} 
\end{figure*}

We consider an FL environment with $K$ clients and a global server, and perform experiments under two schemes on both MLP and SNN-MLP networks: (i) the classical FL \cite{fedavg}, and (ii) the proposed FedAT. The classical FL uses the original data distributed across the $K$ clients, whereas the FedAT employs AT-driven synthetic data generation at the local clients to reduce the adverse effects of class imbalance. We use the FedAvg for the model aggregation for both the classical FL and our proposed FedAT. Note that the evaluation also involves using varying communication rounds to study the convergence capability of the model. The quantitative results for the classical FL and our FedAT are given in Table \ref{tab:fl_results}, and Table \ref{tab:fedat_results}, respectively. These experimental results show that both FL schemes produce stable results for $T$ = 60 communication rounds. In particular, FedAT outperforms the classical FL in terms of all the $P$ (68.16\%, and 74.21\% on the CERT v4.2 and CERT v5.2 datasets, respectively are achieved by our FedAT while 64.05\%, and 51.09\% on the CERT v4.2 and CERT v5.2 datasets, respectively by the classical FL), $R$ (67.74\%, and 73.68\% on the CERT v4.2 and CERT v5.2 datasets, respectively are achieved by our FedAT while 64.00\%, and 51.25\% on the CERT v4.2 and CERT v5.2 datasets, respectively by the classical FL), and $F$ (67.95\%, and 73.94\% on the CERT v4.2 and CERT v5.2 datasets, respectively are achieved by our FedAT while 64.02\%, and 51.17\% on the CERT v4.2 and CERT v5.2 datasets, respectively by the classical FL) metrics. In addition, SNN-MLP proves to be more efficient in FedAT than other combinations of methods for both the CERT v4.2 and v5.2 datasets. Though AT in the centralized ML scheme has a slightly higher performance, the FedAT performs almost equally well in nearly all cases. These results, therefore, clearly show the effectiveness of the FedAT-enabled distributed ITD scheme. In addition, Fig. \ref{fig:mlp_metrics} and Fig. \ref{fig:snnmlp_metrics} depict the performance metrics and the training loss during each communication round in the classical FL using MLP and the proposed FedAT using SNN-MLP, respectively. To summarize, the experiments have shown that by using collaborative training with our proposed FedAT, it is possible to achieve significant results while maintaining the data privacy of the different clients for distributed ITD. As the effective selection of hyperparameters significantly impacts the performance of the FL approaches, we perform experiments with varying numbers of clients, local epochs, and local batch sizes in Section \ref{extended_experiments} to investigate the performance of the proposed FedAT model under various hyperparameter configurations and identify the appropriate parameters over the class-level heterogeneity. 
\begin{table*}[!hbt]
   	\centering
   	\caption{\textcolor{black}{Performance of the classical FL.}}
   	\label{tab:fl_results}
   	\begin{tabular}{ccccccccccc}
   		\hline
   		\multirow{3}{*}{Dataset}    & \multirow{3}{*}{Method} & \multicolumn{9}{c}{FL $(B= 128, E= 1, K=3,$ and $\eta = 0.001)$}                                                             \\ \cline{3-11} 
   		&                         & \multicolumn{3}{c}{$T$ = 10} & \multicolumn{3}{c}{$T$ = 30} & \multicolumn{3}{c}{$T$=60} \\ \cline{3-11} 
   		&                         & $P$       & $R$       & $F$      & $P$       & $R$       & $F$      & $P$      & $R$      & $F$      \\ \hline
   		\multirow{2}{*}{CERT   4.2} & MLP                     & 0.2496  & 0.2492  & 0.2494 & 0.3923  & 0.3761  & 0.3840 & 0.4392 & 0.4384 & 0.4388 \\ 
   		& SNN-MLP                 & 0.4295  & 0.4254  & 0.4274 & 0.5732  & 0.5408  & 0.5565 & 0.6405 & 0.6400 & 0.6402 \\ \hline
   		\multirow{2}{*}{CERT   5.2} & MLP                     & 0.2493  & 0.2496  & 0.2494 & 0.3683  & 0.3546  & 0.3613 & 0.4466 & 0.4407 & 0.4436 \\  
   		& SNN-MLP                 & 0.3266  & 0.3116  & 0.3189 & 0.4190  & 0.4155  & 0.4172 & 0.5109 & 0.5125 & 0.5117 \\ \hline
   	\end{tabular}
\end{table*}

\begin{table*}[!hbt]
   	\centering
   	\caption{\textcolor{black}{Performance of the proposed FedAT.}}
   	\label{tab:fedat_results}
   	\begin{tabular}{ccccccccccc}
   		\hline
   		\multirow{3}{*}{Dataset}    & \multirow{3}{*}{Method} & \multicolumn{9}{c}{Proposed FedAT $(B= 128, E= 1, K=3,$ and $\eta = 0.001)$}                                                            \\ \cline{3-11} 
   		&                         & \multicolumn{3}{c}{$T$ = 10} & \multicolumn{3}{c}{$T$ = 30} & \multicolumn{3}{c}{$T$=60} \\ \cline{3-11} 
   		&                         & $P$       & $R$       & $F$      & $P$       & $R$       & $F$      & $P$      & $R$      & $F$      \\ \hline
   		\multirow{2}{*}{CERT   4.2} & MLP                     & 0.3547  & 0.3761  & 0.3651 & 0.4004  & 0.4044  & 0.4024 & 0.5898 & 0.5825 & 0.5861 \\ 
   		& SNN-MLP                 & 0.5518  & 0.5487  & 0.5502 & 0.6201  & 0.6210  & 0.6206 & 0.6816 & 0.6774 & 0.6795 \\ \hline
   		\multirow{2}{*}{CERT   5.2} & MLP                     & 0.4138  & 0.4065  & 0.4101 & 0.5315  & 0.5133  & 0.5222 & 0.6220 & 0.6263 & 0.6241 \\ 
   		& SNN-MLP                 & 0.6796  & 0.6427  & 0.6606 & 0.6951  & 0.6807  & 0.6878 & 0.7421 & 0.7368 & 0.7394 \\ \hline
   	\end{tabular}
\end{table*}

\begin{table*}[!hbt]
   	\caption{\textcolor{black}{Extended performance analysis of the proposed FedAT on different parameters and FL optimization methods ($T=60, B=128, E = 1,$ and $\eta=0.001$).}}
   	\label{tab:results_b128}
   	\begin{tabular}{clllllllllllll}
   		\hline
   		\multirow{4}{*}{\begin{tabular}[c]{@{}c@{}}$K$\end{tabular}} & \multicolumn{1}{c}{\multirow{4}{*}{Model}} & \multicolumn{12}{c}{FedAT}                                                                                                                                                                                                                                           \\ \cline{3-14} 
   		& \multicolumn{1}{c}{}                       & \multicolumn{6}{c}{CERT   v4.2}                                                                                                               & \multicolumn{6}{c}{CERT v5.2}                                                                                                                 \\ \cline{3-14} 
   		& \multicolumn{1}{c}{}                       & \multicolumn{3}{c}{FedAvg}                                            & \multicolumn{3}{c}{FedProx}                                           & \multicolumn{3}{c}{FedAvg}                                            & \multicolumn{3}{c}{FedProx}                                           \\ \cline{3-14} 
   		& \multicolumn{1}{c}{}                       & \multicolumn{1}{c}{$P$} & \multicolumn{1}{c}{$R$} & \multicolumn{1}{c}{$F$} & \multicolumn{1}{c}{$P$} & \multicolumn{1}{c}{$R$} & \multicolumn{1}{c}{$F$} & \multicolumn{1}{c}{$P$} & \multicolumn{1}{c}{$R$} & \multicolumn{1}{c}{$F$} & \multicolumn{1}{c}{$P$} & \multicolumn{1}{c}{$R$} & \multicolumn{1}{c}{$F$} \\ \hline
   		\multirow{2}{*}{3}                                                             & MLP                                        & 0.6571                & 0.6349                & 0.6458                & 0.6844                & 0.6842                & 0.6843                & 0.5882                & 0.5965                & 0.5923                & 0.5846                & 0.5867                & 0.5856                \\  
   		& SNN - MLP                                  & 0.7433                & 0.7403                & 0.7418                & 0.7432                & 0.7271                &                       & 0.7464                & 0.7499                & 0.7481                & 0.7421                & 0.7419                & 0.7420                \\ \hline
   		\multirow{2}{*}{5}                                                             & MLP                                        & 0.6133                & 0.6809                & 0.6453                & 0.6201                & 0.6171                & 0.6186                & 0.6615                & 0.6452                & 0.6532                & 0.6745                & 0.6758                & 0.6751                \\ 
   		& SNN - MLP                                  & 0.7064                & 0.6911                & 0.6987                & 0.7196                & 0.7061                & 0.7128                & 0.6712                & 0.6758                & 0.6735                & 0.6809                & 0.6384                & 0.6590                \\ \hline
   	\end{tabular}
\end{table*}

\begin{table*}[!hbt]
   	\caption{\textcolor{black}{Extended performance analysis of the proposed FedAT on different parameters and FL optimization methods ($T=60, B=256, E = 1,$ and $\eta=0.001$).}}
   	\label{tab:results_b256}
   	\begin{tabular}{clllllllllllll}
   		\hline
   		\multirow{4}{*}{\begin{tabular}[c]{@{}c@{}}$K$\end{tabular}} & \multicolumn{1}{c}{\multirow{4}{*}{Model}} & \multicolumn{12}{c}{FedAT}                                                                                                                                                                                                                                             \\ \cline{3-14} 
   		& \multicolumn{1}{c}{}                       & \multicolumn{6}{c}{CERT   v4.2}                                                                                                               & \multicolumn{6}{c}{CERT v5.2}                                                                                                                 \\ \cline{3-14} 
   		& \multicolumn{1}{c}{}                       & \multicolumn{1}{c}{$P$} & \multicolumn{1}{c}{$R$} & \multicolumn{1}{c}{$F$} & \multicolumn{1}{c}{$P$} & \multicolumn{1}{c}{$R$} & \multicolumn{1}{c}{$F$} & \multicolumn{1}{c}{$P$} & \multicolumn{1}{c}{$R$} & \multicolumn{1}{c}{$F$} & \multicolumn{1}{c}{$P$} & \multicolumn{1}{c}{$R$} & \multicolumn{1}{c}{$F$} \\ \cline{3-14} 
   		& \multicolumn{1}{c}{}                       & \multicolumn{3}{c}{FedAvg}                                            & \multicolumn{3}{c}{FedProx}                                           & \multicolumn{3}{c}{FedAvg}                                            & \multicolumn{3}{c}{FedProx}                                           \\ \hline
   		\multirow{2}{*}{3}                                                              & MLP                                        & 0.5899                & 0.5486                & 0.5685                & 0.6075                & 0.6033                & 0.6054                & 0.5576                & 0.5524                & 0.5550                & 0.6068                & 0.5965                & 0.6016                \\ 
   		& SNN - MLP                                  & 0.6712                & 0.6686                & 0.6699                & 0.6816                & 0.6417                & 0.6610                & 0.6263                & 0.6246                & 0.6255                & 0.6374                & 0.6328                & 0.6351                \\ \hline
   		\multirow{2}{*}{5}                                                              & MLP                                        & 0.5091                & 0.5148                & 0.5120                & 0.5266                & 0.5211                & 0.5238                & 0.5431                & 0.5408                & 0.5419                & 0.5539                & 0.5407                & 0.5472                \\ 
   		& SNN - MLP                                  & 0.6143                & 0.6089                & 0.6116                & 0.6221                & 0.6192                & 0.6206                & 0.6613                & 0.6405                & 0.6507                & 0.6744                & 0.6677                & 0.6710                \\ \hline
   	\end{tabular}
\end{table*}

\begin{figure*}[!t]
   	\centering
   	\subfloat[][FedAT using $K=3, B=128,$ and $ E=1$.]{\includegraphics[width=0.4\textwidth]
   		{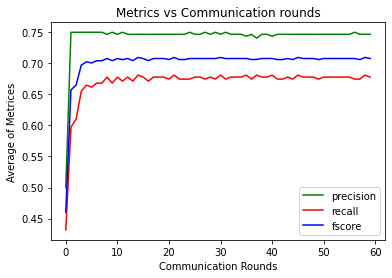}
   		\label{fig:b128e1}}
   	\subfloat[][FedAT using $K=3, B=128,$ and $ E=5$.]{\includegraphics[width=0.4\textwidth]
   		{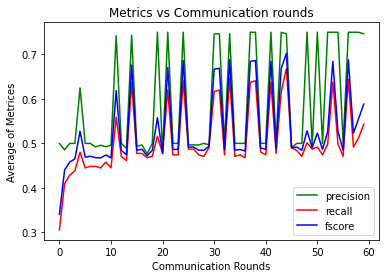}
   		\label{fig:b128e5}}
   	\caption{Effect of the local epochs on FedAT. }  
   	\label{fig:b128e15_main} 
\end{figure*}

\subsubsection{Generator Architecture}
We use the same ACGAN architecture for centralized ML, classical FL, and our FedAT experiments. The G model is designed with $\{32, 64, 128\}$ neurons. The G model uses the Gaussian noise and the latent dimension as inputs. The latent dimension provided as input to the G model is equal to the number of features in the dataset. The GAN training is performed for 200 epochs. Once the GAN training is over, the model generates synthetic samples of the insider scenarios resent in the client. We consider the improvement in the performance metrics as the criteria for validating the similarity of the synthetic samples. Since the data generated from the client locations could mimic the original distribution, the FL approaches produce quality results, as seen in  Table \ref{tab:fl_results}, and Table \ref{tab:fedat_results}.

\subsubsection{How Does SNN Help in Training?} 
The performance improvement of using the SNN-MLP is evident from the results, as shown in Table \ref{tab:benchmark}, Table \ref{tab:fl_results}, and Table \ref{tab:fedat_results}. The intuition behind the self-normalization in SNN is to maintain the mean and variance of each neural network layer near 0 and 1. Incorporating the SELU activation function, the self-normalizing layer has been demonstrated to produce three unique features. The network's average learning rate is controlled by the SELU activation function using positive and negative values. With the help of a continuous curve, the SELU keeps a fixed point in the neural network. The SELU activation function guarantees a zero mean and one variance for the activation output. This enables it to train deeper neural networks without incurring substantial gradient degradation. These characteristics contribute to minimizing the model's invariance. 


\subsubsection{Extended Experiments} \label{extended_experiments}
In this part, we study the impact of using different hyperparameters on the FedAT scheme. This experiment comprises varying clients ($K$), batch size ($B$), and local epochs ($E$). In addition, we employ the FedProx FL optimization method \cite{fedprox}, an improved version of FedAvg on non-IID data, for comparison as well. The quantitative results are shown in Table \ref{tab:results_b128} and Table \ref{tab:results_b256} (on varying $K$ and $B$), and the qualitative results in Fig. \ref{fig:b128e15_main} (on varying $E$). For both datasets, the tables manifest that our FedAT performs slightly better with smaller batch sizes i.e., $B=128$ on any number of clients using both federated optimization schemes (FedAvg and FedProx). 
On the other hand, Fig. \ref{fig:b128e1} shows a stable convergence in FedAT when the local training is performed with a single local epoch, whereas Fig. \ref{fig:b128e5} shows a fluctuation in the performance metrics using $E=5$ in the local training. Overall, these results suggest to choosing a smaller $B$ and $E$ in the FedAT approach.

\section{Conclusion and Future Work}\label{sec_conclusion}
In this paper, we have proposed a practical approach for using the FL and AT paradigms in the context of distributed ITD. In particular, we have discussed the need for an FL-enabled solution approach for ITD and GAN-based AT to help reduce the class imbalance in the non-IID dataset of different clients. In addition, our FedAT-based scheme helps build an effective ITD model for all the clients involved in deep cooperative learning without sharing their confidential employee information. We have provided experiments using two public datasets (CERT v4.2 and CERT v5.2), and two federated optimization methods (FedAvg and FedProx), which manifest the supervisor performance of our FedAT scheme over the classical FL strategies. In the future, we plan to redesign our FedAT architecture using heterogeneous learning models for different clients. Furthermore, we would focus on improving the AT model by incorporating mechanisms that can defend against the backdoor and model inversion attacks in FL.

\bibliographystyle{ieeetr}

\begin{thebibliography}{00}
	
\bibitem{insider_survey}
Homoliak, I., Toffalini, F., Guarnizo, J., Elovici, Y. and Ochoa, M., 2019. Insight into insiders and it: A survey of insider threat taxonomies, analysis, modeling, and countermeasures. ACM Computing Surveys (CSUR), 52(2), pp.1-40.
	
\bibitem{fed_learning2}
Yang Q., Liu Y., Chen T., Tong Y.Federated machine learning: Concept and applications ACM Trans. Intell. Syst. Technol. (TIST), 10 (2) (2019), pp. 1-19


\bibitem{fl_nlp}
Ramaswamy, S.; Mathews, R.; Rao, K.; Beaufays, F. Federated Learning for Emoji Prediction in a Mobile Keyboard. arXiv 2019,
arXiv:1906.04329

\bibitem{fl_iot}
Zhang, L., Xu, J., Vijayakumar, P., Sharma, P.K. and Ghosh, U., 2022. Homomorphic Encryption-based Privacy-preserving Federated Learning in IoT-enabled Healthcare System. IEEE Transactions on Network Science and Engineering.

\bibitem{fl_healthcare}
Nguyen, D.C., Pham, Q.V., Pathirana, P.N., Ding, M., Seneviratne, A., Lin, Z., Dobre, O. and Hwang, W.J., 2022. Federated learning for smart healthcare: A survey. ACM Computing Surveys (CSUR), 55(3), pp.1-37.

\bibitem{dl_data_shortage}
Sun, C., Shrivastava, A., Singh, S. and Gupta, A., 2017. Revisiting unreasonable effectiveness of data in deep learning era. In Proceedings of the IEEE international conference on computer vision (pp. 843-852).


\bibitem{insider1}
Chattopadhyay, P., Wang, L. and Tan, Y.P., 2018. Scenario-based insider threat detection from cyber activities. IEEE Transactions on Computational Social Systems, 5(3), pp.660-675.

\bibitem{insider2}
Le, D.C. and Zincir-Heywood, N., 2021. Anomaly detection for insider threats using unsupervised ensembles. IEEE Transactions on Network and Service Management, 18(2), pp.1152-1164.

\bibitem{insider3}
Le, D.C., Zincir-Heywood, N. and Heywood, M.I., 2020. Analyzing data granularity levels for insider threat detection using machine learning. IEEE Transactions on Network and Service Management, 17(1), pp.30-44.

\bibitem{insider4}
Yuan, S. and Wu, X., 2021. Deep learning for insider threat detection: Review, challenges and opportunities. Computers \& Security, 104, p.102221.


\bibitem{fedavg}
McMahan, B., Moore, E., Ramage, D., Hampson, S. and y Arcas, B.A., 2017, April. Communication-efficient learning of deep networks from decentralized data. In Artificial intelligence and statistics (pp. 1273-1282). PMLR.

\bibitem{fedprox}
Tian Li, Anit Kumar Sahu, Manzil Zaheer, Maziar Sanjabi, Ameet Talwalkar, and
Virginia Smith. 2018. Federated optimization in heterogeneous networks. arXiv
preprint arXiv:1812.06127 (2018).

\bibitem{adv_samples}
Jiliang Zhang and Chen Li. 2019. Adversarial examples: Opportunities and
challenges. IEEE transactions on neural networks and learning systems 31, 7 (2019), 2578–2593.

\bibitem{adv_data_aug}
Riccardo Volpi, Hongseok Namkoong, Ozan Sener, John Duchi, Vittorio Murino,
and Silvio Savarese. 2018. Generalizing to unseen domains via adversarial data
augmentation. arXiv preprint arXiv:1805.12018 (2018).

\bibitem{adv_samples_1}
Andrew Ilyas, Shibani Santurkar, Dimitris Tsipras, Logan Engstrom, Brandon
Tran, and Aleksander Madry. 2019. Adversarial examples are not bugs, they are
features. arXiv preprint arXiv:1905.02175 (2019).

\bibitem{gan}
Goodfellow I., Pouget-Abadie J., Mirza M., Xu B., Warde-Farley D., Ozair S., Courville A., Bengio Y.Generative adversarial nets. Advances in Neural Information Processing Systems (2014), pp. 2672-2680

\bibitem{cgan}
Mirza M., Osindero S.Conditional generative adversarial nets(2014)
ArXiv Preprint arXiv:1411.1784

\bibitem{acgan}
Odena, A., Olah, C. and Shlens, J., 2017, July. Conditional image synthesis with auxiliary classifier gans. In International conference on machine learning (pp. 2642-2651). PMLR.

\bibitem{fl_adv_learn}
Xingchao Peng, Zijun Huang, Yizhe Zhu, and Kate Saenko. 2019. Federated
adversarial domain adaptation. arXiv preprint arXiv:1911.02054 (2019).

\bibitem{fl_ids_survey}
Agrawal, S., Sarkar, S., Aouedi, O., Yenduri, G., Piamrat, K., Alazab, M., Bhattacharya, S., Maddikunta, P.K.R. and Gadekallu, T.R., 2022. Federated learning for intrusion detection system: Concepts, challenges and future directions. Computer Communications.

\bibitem{insider_cgan_ours}
Gayathri, R.G., Sajjanhar, A., Xiang, Y. and Ma, X., 2021, October. Anomaly detection for scenario-based insider activities using cgan augmented data. In 2021 IEEE 20th International Conference on Trust, Security and Privacy in Computing and Communications (TrustCom) (pp. 718-725). IEEE.


\bibitem{priv_fl_gan}
Private fl-gan: Differential privacy synthetic data generation based on federated learning

\bibitem{mdgan}
Hardy, C., Le Merrer, E. and Sericola, B., 2019, May. Md-gan: Multi-discriminator generative adversarial networks for distributed datasets. In 2019 IEEE international parallel and distributed processing symposium (IPDPS) (pp. 866-877). IEEE.


\bibitem{snn}
Klambauer, G., Unterthiner, T., Mayr, A. and Hochreiter, S., 2017. Self-normalizing neural networks. Advances in neural information processing systems, 30.

\bibitem{snn_vuln}
J.-H. Jacobsen, J. Behrmann, R. Zemel, M. Bethge
Excessive invariance causes adversarial vulnerability
International Conference on Learning Representations (2018)


\bibitem{lecun_layer}
Y.A. LeCun, L. Bottou, G.B. Orr, K.-R. Müller
Efficient Backprop Neural networks: Tricks of the trade, Springer (2012), pp. 9-48

\bibitem{cmu}
C. M. U. CERT Team, CMU CERT synthetic insider threat data sets. Available at : https://resources.sei.cmu.edu/library/asset-view.cfm?assetid=508099.

\end{thebibliography}
\end{document}